# Universal mechanism of bandgap engineering in transition-metal dichalcogenides

*Mingu Kang,[†] Beomyoung Kim,[†,‡] Sae Hee Ryu,[†] Sung Won Jung,[†] Jimin Kim,[†] Luca Moreschini,[†,‡] Chris Jozwiak,[‡] Eli Rotenberg,[‡] Aaron Bostwick,[‡] and Keun Su Kim*[†]*

[†]Department of Physics, Pohang University of Science and Technology, Pohang 37673, Korea

[‡]Advanced Light Source, E. O. Lawrence Berkeley National Laboratory, Berkeley, CA 94720, USA

*E-mail: keunsukim@postech.edu.

ABSTRACT. Two-dimensional (2D) van-der-Waals semiconductors have emerged as a class of materials with promising device characteristics owing to the intrinsic bandgap. For realistic applications, the ideal is to modify the bandgap in a controlled manner by a mechanism that can be generally applied to this class of materials. Here, we report the observation of a universally tunable bandgap in the family of bulk 2*H* transition metal dichalcogenides (TMDs) by *in situ* surface doping of Rb atoms. A series of angle-resolved photoemission spectra unexceptionally shows that the bandgap of TMDs at the zone corners is modulated in the range of 0.8 ~ 2.0 eV, which covers a wide spectral range from visible to near infrared, with a tendency from indirect to direct bandgap. A key clue to understand the mechanism of this bandgap engineering is provided by the spectroscopic signature of symmetry breaking and resultant spin splitting, which can be explained by the formation of 2D electric dipole layers within the surface bilayer of TMDs. Our results establish the surface Stark effect as a universal mechanism of bandgap engineering based on the strong 2D nature of van-der-Waals semiconductors.

KEYWORDS: Bandgap engineering, two-dimensional semiconductors, giant Stark effect, transition-metal dichalcogenides

The bandgap is a fundamental property of semiconducting materials that determines their electronic and optical properties.[1] As such, it not only plays a central role in the working principle of modern semiconductor devices, but also sets the limit of device performances. Bandgap engineering is a powerful technique to overcome the limit of natural properties and also to design novel materials properties and functionalities.[2,3] In conventional semiconductors, however, bandgap engineering has been achieved by the variation of chemical strain or layered heterostructures, which all requires heavy structural modifications. The key issue is thus to develop a universal mechanism of bandgap engineering that is electrically tunable and easily viable in gated devices.

In recent years, 2D van-der-Waals crystals, such as graphene, black phosphorus, and TMDs, has emerged as a class of materials, in which their strong 2D nature offers unprecedented opportunities of bandgap engineering.[4-7] The 2$H$ phase of TMDs (WS$_2$, MoS$_2$, WSe$_2$, MoSe$_2$, and MoTe$_2$) is a layered semiconductor with great potential for application in nanoelectronic and optoelectronic devices.[8] Their single layer consists of transition-metal atoms (Mo or W) and chalcogen atoms (S, Se, or Te), arranged into the hexagonal honeycomb network similar to graphene, as shown in Figure 1a. In their hexagonal Brillouin zone (Figure 1b), the lack of inversion symmetry and spin-orbit coupling affects valence bands of transition-metal $d_{x2-y2}$ and $d_{xy}$ orbitals around zone corners, leading to the out-of-plain spin polarization reversed between two different valleys (K and K').[9,10] For multilayers, two single-layers are stacked as a unit bilayer by weak van-der-Waals forces, recovering inversion symmetry (centred at small black circles in Figure 1a,b). This guarantees valence bands localized in different layers (L1 and L2) to have opposite spin splitting and thus doubly degenerate,[9-12] as illustrated in Figure 1b. The resulting low-energy band structure (Figure 1c,d) shows valence bands (VB, red) and conduction bands

(CB, yellow) that are separated by an indirect gap (arrow in Figure 1d) in the range of 1.0 ~ 1.4 eV (ref. 8)

The band structure of multilayer 2*H*-TMDs is highly susceptible to various physical parameters, such as thickness,[13] stacking,[14] strain,[15,16] and electric field.[17,18] Although the structural variation of thickness and stacking modulates the magnitude of the bandgap, there is little change in the local gap at the K point,[8] which is critical for optical properties of 2*H*-TMDs.[19,20] On the other hand, the application of external electric field vertically to TMD layers, which is more viable in gated devices, is widely predicted to modulate the local bandgap at the K point.[17,18] Since this mechanism in principle works for TMDs down to the bilayer limit regardless of their thickness (Supporting Figure S1), we chose bilayer 2*H*-TMDs as a minimal model to capture essential features of how the band structure changes with vertical electric field (Figure 1d–f). The electronic states confined in TMD bilayers under electric field monotonically increase the energy of VB by the negative potential, and decreases the energy of CB by the positive potential.[17,18] The formation of electric dipole-like layers effectively reduces the magnitude of the bandgap, which is often termed the giant Stark effect in the study of nanoribbons and nanotubes.[21–23] The field-induced energy shifts are different for VBs between Γ and K points, and for CBs between T and K points depending on their orbital characters, which forms a tendency of the indirect-to-direct bandgap transition (Figure 1d–f). Furthermore, the interlayer potential difference breaks inversion symmetry, leading to spin splitting of VBs at around the K point (Figure 1b,f).

Here, we report the realization of electrically tunable bandgap by applying the Stark effect to the surface of bulk 2*H*-TMDs, where the strong 2D van-der-Waals nature enables quantum confinement as in the bilayer of 2*H*-TMDs under vertical electric field. In our experiments, the vertical electric field was generated by chemical surface doping, the *in situ* deposition of alkali-metal atoms, which is proven effective on the surface of 2D van-der-Waals crystals.[4,6,24] The Rb atoms distributed arbitrarily on the surface donate charges to adjacent TMD layers. The ionized Rb atoms produces strong vertical electric field, leading to the Stark effect and resultant bandgap changes, which are directly measured by angle-resolved photoemission spectroscopy (ARPES).

In Figure 2, we display a doping series of ARPES spectra, taken from five different 2*H*-TMDs marked on the left. As expected for pristine 2*H*-TMDs (red lines in Figure 1d), there is commonly a well-defined VB in Figure 2a,e,i,m,q. Around the K point, two branches of VB originate from layer-localized spin polarizations discussed in Figure 1b, which gradually merge into a single branch of out-of-plain $d_{z^2}$ orbitals at around the time-reversal invariant Γ point. The VB maximum is located at the Γ point, and its energy scale with respect to the Fermi energy ($E_F$) is greater than 1.0 eV. Considering the well-known bandgap of bulk TMDs (1.0 ~ 1.4 eV),[8,25-27] we found that our near-neutral samples are n-doped near the surface with band bending towards the bulk. Based on the Thomas-Fermi screening theory, we estimate the screening length of space-charge layers to be about 280 nm (Supporting Information). This built-in electric field spread over a few hundred nanometers is extremely weak at the atomic scale, so that the spin-orbit doublet of VBs at the K point remains nearly degenerated in Figure 2a,e,i,m,q.

Upon the deposition of Rb atoms, the donated electrons populate the CB of TMDs, and its minimum at the K and/or T point appears below $E_F$ in ARPES (Figure 2b,f,j,n,r). From this point, a direct or an indirect bandgap as well as its magnitude can be directly measured by ARPES. As increasing the dopant density, we found that the centre energy (red dotted lines in Figure 2a–d) between the VB maximum and the CB minimum gradually shifts down towards higher energies, indicating steady doping. Furthermore, with respect to this centre energy, VB and CB get progressively closer to each other, which is due to strong vertical electric field generated from the ionized Rb atoms on the surface of 2$H$-TMDs. As a result, the magnitude of the bandgap gradually reduces with the strength of electric field, which is in excellent agreement with our theoretical band calculations in Figure 1e,f. The key aspects of our observations, electron doping as well as bandgap reductions, are unexceptionally observed in the whole family of semiconducting 2$H$-TMDs that we measured in this study (Figure 2).

The bandgap modulation can be more clearly identified in a continuous doping series of ARPES spectra, and its 3D representation is shown for WS$_2$ in Figure 3a (the others show essentially the identical behaviour, see Supporting Figure S2). The constant-energy cut at $E_F$ shows that the Fermi momentum ($k$) of CBs at both K and T points steadily increases with the deposition time. The electron concentration $n$ estimated based on Luttinger's theorem is found to be linearly proportional to the dopant density (Supporting Figure S3), indicating a monotonic charge transfer. The constant $k$ cut at the K point shows the energy variation of the CB minimum $E_c$ and the VB maximum $E_v$ with the deposition time. At first, $E_v$ rapidly shifts down by the dominant doping effect, and then gradually turns up, as compensated by the Stark effect. From $E_c - E_v$, the magnitude of the direct gap at the K point, which is critical for optical properties of 2$H$-TMDs,[19,20] is plotted as a function of $n$ in Figure 3b. Consequently, the bandgap at the K point is modulated in the range of 0.8 ~ 2.0 eV, covering a wide spectral range from visible to near infrared.

The energy shift of VB and CB with $n$ can be different for $k$, depending on their orbital characters. The energy shift of CB at the K point is slightly larger than that of CB at the T point (Figure 2i–l and Figure 3c for WSe$_2$), making a crossover at $n = 5.9 \times 10^{13}$ cm$^{-2}$ (dotted line in Figure 3c). At the high $n$ limit, $E_c$ occurs at the K point for all 2$H$-TMDs (Figure 2d,h,l,p,t). On the other hand, the energy shift of VB at the K point is greater than that of VB at the Γ point (Figure 3d). This together forms a tendency of the transition from indirect to direct bandgap at the K point consistently for 2*H*-TMDs, and the actual transition is observed for MoTe$_2$ (Figure 2t). Therefore, a direct bandgap, which is a key feature of monolayer TMDs for their unusual optical properties,[19,20] can selectively be induced by the electric-field effect in the surface of bulk TMDs or thin films. This field-induced indirect-to-direct bandgap transition would more easily be achieved by applying the same mechanism to the bilayer of MoSe$_2$, WSe$_2$, and their heterostructures,[28] where the energy difference of VBs between Γ and K points is relatively small in their pristine states.

An important clue to understand the microscopic mechanism of bandgap engineering can be obtained from the layer-localized spin polarization of VBs around the K point (Figure 2a,e,i,m,q). Upon the deposition of dopants, the doublet of spin-layer locking splits into a pair of doublets, which are most clearly observed for WS$_2$ (Figure 2d and Figure 3a) and consistent with recent observations for WSe$_2$ (ref. 24). This is a direct spectroscopic signature of lifting spin degeneracy by breaking interlayer symmetry in the unit bilayer of 2*H*-TMDs.[11,12] We found that the energy of spin splitting increases monotonously with $n$ up to 0.2 eV commonly for five 2*H*-TMDs (Figure 3e). As expected for TMD bilayers (Figure 1e,f), the effect of strong vertical electric field decreases the energy of VBs localized in the upper layer (denoted as L1), and increases that of VBs localized in the bottom layers (denoted as L2). Indeed, the ARPES intensity of L1, averaged over finite $k$ and photon energies to eliminate the matrix-element effect, is always greater than that of L2 due to the finite escape depth (Supporting Figure S4). Furthermore, the relative degree of spin splitting among 2*H*-TMDs (Figure 3e) shows a positive correlation with bandgap modulations in Figure 3b (see Supporting Table S1), confirming the proposed Stark mechanism.

In Figure 4a, we plot the energy shift of L1 and L2 relative to that of layer-delocalized VB at the Γ point (Figure 3d). The potential variations (Δ) of L1 and L2 sublayers in the surface bilayer are opposite to each other (negative and positive, respectively) with their magnitude ratio $\Delta_{L2}/\Delta_{L1}$ about 0.7. This is evidence of 2D electric dipole layers within the surface bilayer of 2$H$-TMDs, screening vertical electric field generated by the adjacent Rb atoms, as illustrated in Figure 4b,c. The formation of 2D dipole layers is consistent with the damped oscillation of relative change densities (inset of Figure 4c) predicted theoretically for field-effect doping.[29] This interfacial potential landscape is distinct from the simplified picture of smooth band bending adopted in conventional semiconductors (dotted line in Figure 4c),[3] resulting from the strong 2D nature of van-der-Waals semiconductors, where electronic states are delocalized along in-plain directions while strongly localized in the out-of-plain direction (reminiscent of atomic-scale parallel capacitors). This replaces the role of quantum confinement required for the Stark mechanism in nanoscale materials, which we call as the surface Stark effect on 2D van-der-Waals semiconductors.

Consequently, we demonstrate the surface Stark effect universally in all available semiconducting 2$H$-TMDs. The similar phenomenology has also been observed in another 2D van-der-Waals semiconductor, black phosphorus,[6] but its microscopic mechanism could be unfolded in this study owing to the unique spin-layer splitting of 2$H$-TMDs. These results together establish the surface Stark effect as a universal mechanism of bandgap engineering in 2D semiconductors. Since 2$H$-TMDs are a layered material in which individual 2D layers are coupled by only weak van-der-Waals force, the surface Stark effect that we demonstrate here in the surface bilayer of bulk 2$H$-TMDs should in principle work in the surface bilayer of few-layer samples or even in bilayer samples, as supported by first-principles and tight-binding band calculations.[17,18] This mechanism would be useful for the design and optimization of field-effect optoelectronic devices made of 2D semiconductors. The surface doping concentration required for this level of bandgap engineering is estimated to be of the order of $10^{13}$ cm$^{-2}$, which is attainable in gated devices[31] or by ionic liquid gating.[11] From the potential difference between L1 and L2, we can roughly estimate the applied electric field at maximum to be 0.17 V/nm for WS$_2$. Finally, we note that the mechanism revealed here should be carefully taken into account in analysing negative electron compressibility in 2$H$-TMDs[24] and related van-der-Waals crystals.

**Experimental Methods.** ARPES experiments were conducted at two different synchrotron radiations, the Beamline 7 of Advanced Light Source (most data) and at the Beamline I05 of Diamond Light Source (part of data in Figure 2). The two ARPES endstations are equipped with R4000 hemispherical electron analysers (VG a, Sweden). Data were collected at 80–90 K (to prevent surface charging) with the photon energy of 45–64 eV for Figure 2 and 80–120 eV for Supporting Figure S4. Energy and momentum resolutions were better than 20 meV and 0.01 Å$^{-1}$, respectively. The single-crystalline 2$H$-TMD samples (HQ graphene) were cleaved in the ultrahigh vacuum chamber with the base pressure better than 4 × 10$^{-11}$ torr. Chemical surface doping, the deposition of alkali-metal atoms on the surface of 2$H$-TMDs, was achieved by *in situ* electrothermal heating of commercial dispensers (SAES). We chose Rb as a dopant to avoid any possible intercalation into the bulk of 2$H$-TMDs.[32] The density of dopants was calibrated in unit of monolayer (ML) from simultaneously taken Rb 4$p$ core-level spectra (Supporting Figure S3).[33] The electron concentration was estimated based on Luttinger's theorem as $n = \Sigma k_F^2/2\pi$, where the summation runs for all the closed Fermi pockets within a single Brillouin zone.

**Theoretical Calculations.** To describe the band structure of bilayer 2$H$-TMDs, we have used the effective tight binding Hamiltonian for monolayer TMD taken from the work by Shanavas *et* *al*. (ref. 34), and extend it to the bilayer case. The bilayer model we presented in Figure 2 is given by

$$H_{2L} = H_{1L} \otimes I_2 + H_{12} + H_E + H_{SO} \qquad (1)$$

where $H_{1L}$, $H_{12}$, $H_E$, and $H_{SO}$ represents monolayer, interlayer hopping, electric field, and spin-orbit coupling terms, respectively. In the monolayer Hamiltonian, we used reduced symmetry by considering the *d* orbital of transition-metal atoms and *s* orbital of two chalcogen atoms. Then, the initial 7 × 7 Hamiltonian can be reduced to 5 × 5 effective Hamiltonian through Löwdin down-folding.[34] To consider the effect of spin-orbit coupling and interlayer coupling, the monolayer Hamiltonian was expanded to the 20 × 20 matrix in spin and layer pseudo-spin space. Then, the interlayer coupling Hamiltonian was added as non-diagonal elements in pseudospin space. In spin-space, the standard spin-orbit matrix element for *d* orbital was used.[34] The effect of vertical electric field was introduced by the on-site potential difference *U* between the bilayers. In each layer, we also introduced the effect of electric field on the on-site energy of initial 7 × 7 Hamiltonian. In the down-folding process, the difference in on-site potential affects hopping parameters, and the band structure changes with the applied electric field. The tight-binding parameters for 2*H*-TMDs were optimized to reproduce their previously reported band structures by first-principles band calculations.[17,18] We find that the critical value for the indirect-to-direct bandgap transition in between Figure 1e,f is about $U_c = 0.15$ eV.

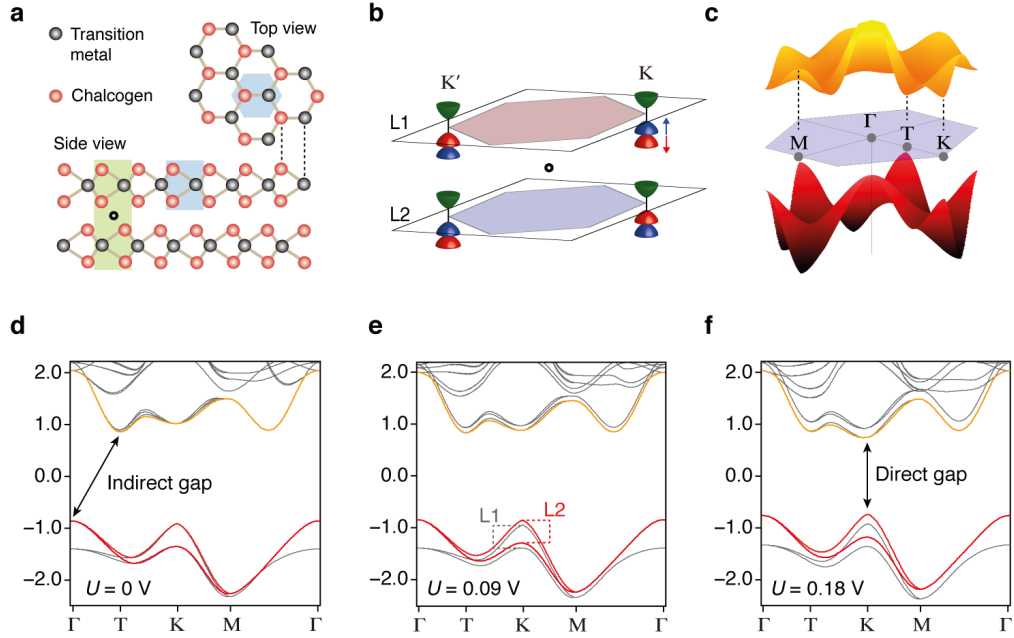

**Figure 1.** Band structure and tunable bandgap of 2*H*-TMDs under electric field. (a) Lattice structure of 2*H*-TMDs. The light blue regions represent the unit cell of monolayer TMDs, whereas the light green region represents the unit cell of bilayer 2*H*-TMDs. (b) Schematics for a low-energy pair of VB (red and blue) and CB (green) around two different valleys (K and K') and in two different layers (L1 and L2). The red and blue arrows indicate out-of-plain spin polarizations. Open circles in a,b denote the inversion center of bilayer 2*H*-TMDs. (c) Low-energy band topology of bilayer 2*H*-TMDs, obtained from tight-binding band calculations. The shaded area between VB (red) and CB (yellow) shows the hexagonal surface Brillouin zone of 2*H*-TMDs with high symmetry points marked by gray dots. (d)−(f), A series of tight-binding band calculations for TMD bilayers with the interlayer potential difference $U$, as marked at the lower left. Double-head arrows in d,f indicate the indirect-to-direct bandgap transition. The splitting of VBs is denoted as L1 and L2, corresponding to their layer origins, as explained in the text.

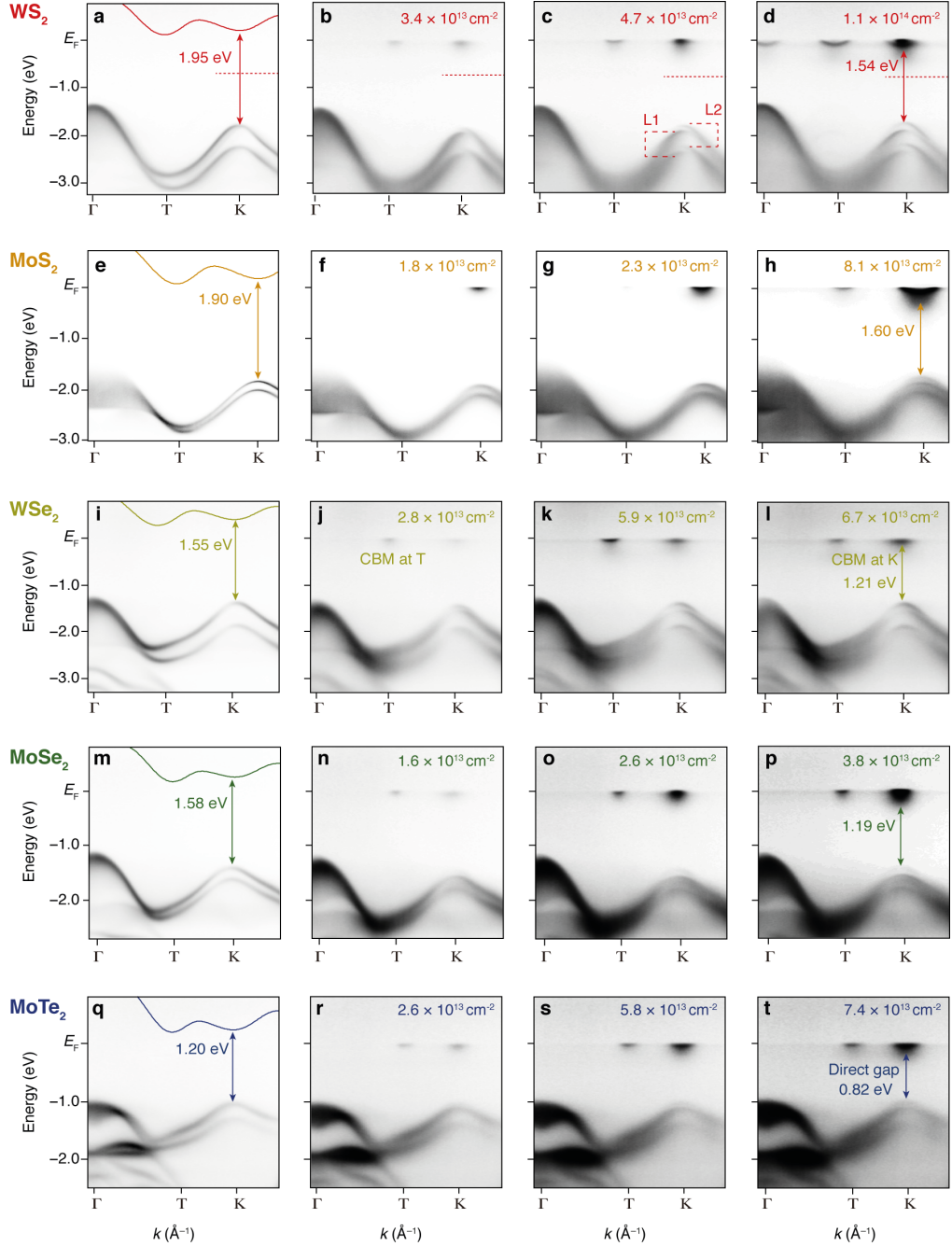

**Figure 2.** Bandstructure evolution of 2*H*-TMDs with surface doping. A doping series of ARPES spectra taken for WS$_2$ (a–d), MoS$_2$ (e–h), WSe$_2$ (i–l), MoSe$_2$ (m–p), and MoTe$_2$ (q–t) along the ΓK direction. The data were collected at 80–90 K with the photon energy of 45–64 eV. Solid lines in a,e,i,m,q show the CB of pristine TMDs, obtained from tight-binding band calculations with the well-known direct bandgap at the K point,[25–27] as indicated by the arrows. Dotted lines in a–d denote the middle point of VB and CB at the K point. After doping, *n* quantified from the Fermi area (Methods) is marked at the upper right of each panel. The splitting of VB is denoted as L1 and L2, as shown in c, indicating their layer origin.

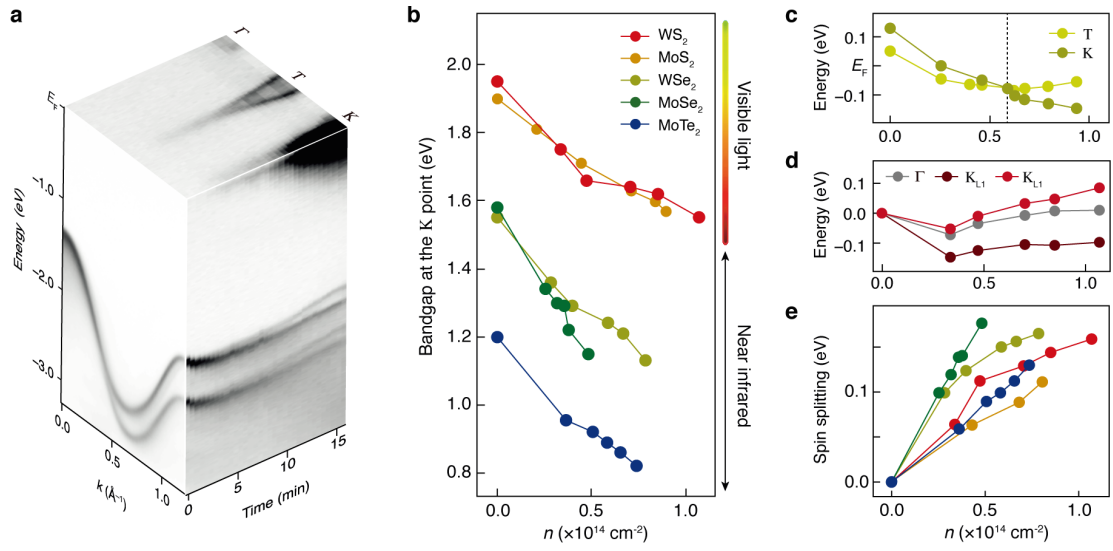

**Figure 3.** Quantitative analysis for the bandgap evolution of 2*H*-TMDs. (a) 3D representation of ARPES spectra taken with fine doping steps for WS$_2$ (for the others, see Supporting Figure S2). The left, right, and top panels show the pristine band dispersion, constant *k* cut at the K point, and constant-energy cut at $E_F$ respectively. (b) Bandgap at the K point plotted as a function of *n* for all 2*H*-TMDs. Shown on the right axis is the spectral range corresponding to the range of tunable bandgaps in the family of 2*H*-TMDs. (c) Energy positions of local CB minima at T and K points for WSe$_2$, and (d) relative energy variations of local VB maxima at Γ and K points for WS$_2$. The dotted line in c indicates the energy crossover of CB minima at T and K points. e, Spin splitting of VB at the K point as a function of *n* for 2*H*-TMDs as marked in b.

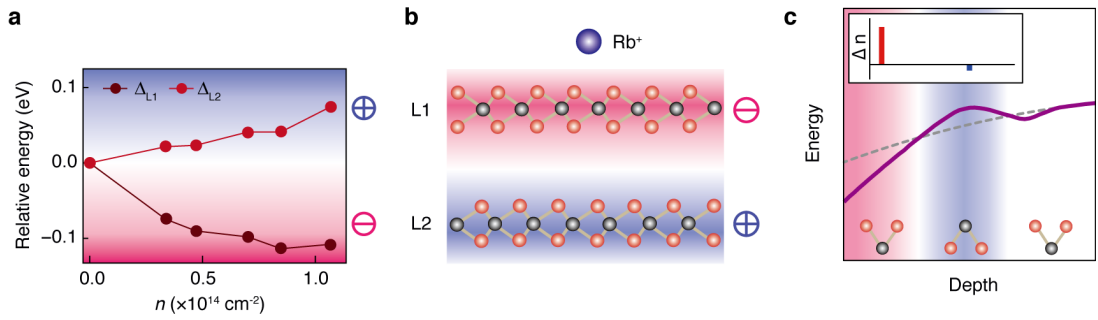

**Figure 4.** Universal mechanism of bandgap engineering. (a) Energy shifts of L1 and L2 relative to that of VB at the Γ point as a function of $n$, taken from Figure 3d. The red and blue regions and markers denote positive and negative potentials, respectively. (b) Schematic illustration of 2D electric dipole layers formed within the surface bilayer of 2$H$-TMDs in response to vertical electric field produced by the ionized Rb atom (dark blue ball). (c) Potential landscape in the surface of 2$H$-TMDs, which is simplified to show the overall trend of energy at transition-metal sites.[30] The gray dotted line denotes the simplified picture of band bending in conventional semiconductors.[3] Inset in c shows the modulation of charge density relative to that of pristine states at transition-metal sites in the surface bilayer of 2$H$-TMDs, taken from first-principles calculations.[29]

## ASSOCIATED CONTENT

**Supporting Information**.

Additional tight-binding calculations for four-layer 2*H*-TMDs; additional ARPES data for continuous doping series, monotonic charge transfer, and layer origin of spin-split bands; table for band parameters; calculation method for the screening length of space charge layers (PDF)

## AUTHOR INFORMATION

**Corresponding Author**

*E-mail: keunsukim@postech.edu.

**Present Addresses**

M.K.: Department of Physics, Massachusetts Institute of Technology, Cambridge, MA 02139, USA.

**Author Contributions**

M.K. conducted ARPES experiments and analysed data with support from B.K., S.H.R., J.K., E.R. and A.B. M.K. performed tight-binding band calculations with help from S.W.J. K.S.K. conceived and directed the project. M.K. and K.S.K. wrote the manuscript with input from all other co-authors.

**Notes**

The authors declare no competing financial interest.

ACKNOWLEDGMENT

This work was supported by the POSTECH Basic Science Research Institute Grant. The Advanced Light Source was supported by the U.S. Department of Energy, Office of Sciences, under contract DE-AC02-05CH11231. We thank Diamond Light Source for access to beamline I05 (proposal number SI13946) that contributed to the results presented here. We thank W. J. Shin, Y. Sohn, M. Huh, T. K. Kim, and M. Hoesch for supporting ARPES experiments.

REFERENCES

(1) Peplow, M. *Nature* **2011,** 503, 327–329.

(2) Capasso, F. *Science* **1987,** 235, 172–176.

(3) Zeghbroeck, B. V. V. *Principles of semiconductor devices and heterojunctions,* Prentice Hall, Upper Saddle River, 2009.

(4) Ohta, T.; Bostwick, A.; Seyller, T.; Horn, K.; Rotenberg, E. *Science* **2006,** 313, 951–954.

(5) Zhang, Y.; Tang, T. T.; Girit, C.; Hao, Z.; Martin, M. C.; Zettl, A.; Crommie, M. F.; Shen, Y. R.; Wang, F. *Nature* **2009,** 459, 820–823.

(6) Kim, J.; Baik, S. S.; Ryu, S. H.; Sohn, Y.; Park, S.; Park, B. G.; Denlinger, J.; Yi, Y.; Choi, H. J.; Kim, K. S. *Science* **2015,** 349, 723–726.

(7) Zeng, Q.; Wang, H.; Fu, W.; Gong, Y.; Zhou, Wu; Ajayan, P. M.; Lou, J.; Liu, Z. *Small* **2015,** 11, 16, 1868–1884.

(8) Wang, Q. H.; Kalantar-Zadeh, K.; Kis, A.; Coleman, J. N; Strano, M. S. *Nat. Nanotechnol.* **2012,** 7, 699–712.

(9) Zhu, Z. Y.; Cheng, C.; Schwingenschlogl, U. *Phys. Rev. B* **2011,** 84, 153402.

(10) Xiao, D.; Liu, G.; Feng, W.; Xu, X.; Yao, W. *Phys. Rev. Lett.* **2012,** 108, 196802.

(11) Yuan, H.; Bahramy, M. S.; Morimoto, K.; Wu, S.; Nomura, K.; Yang, B. J.; Shimotani, H.; Suzuki, R.; Toh, M.; Kloc, C.; Xu, X.; Arita, R.; Nagaosa, N.; Iwasa, Y. *Nat. Phys.* **2013,** 9, 563–569.17

(12) Riley, J. M.; Dendzik, D.; Michiardi, M.; Takayama, T.; Bawden, L.; Granerod, C.; Leandersson, M.; Balasubramanian, T.; Hoesch, M.; Kim, T. K.; Meevasana, W.; Hofmann, Ph.; Bahramy, M. S.; Wells, J. W.; King, P. D. C. *Nat. Phys.* **2014,** 10, 835–839.

(13) Zhang, Y.; Chang, T. R.; Zhou, B.; Cui, Y. T.; Yan, H.; Liu, Z.; Schmitt, F.; Lee, J.; Moore, R.; Chen, Y.; Lin, H.; Jeng, H. T.; Mo, S. K.; Hussain, Z.; Bansil, A.; Shen, Z. X. *Nat. Nanotechnol.* **2014,** 9, 111–115.

(14) Jiang. T.; Liu, H.; Huang, D.; Zhang, S.; Li, Y.; Gong, X.; Shen, Y. R.; Liu, W. T.; Wu, S. *Nat. Nanotechnol.* **2014,** 9, 825–829.

(15) Conley, H. J.; Ziegler, J. I.; Haglund, R. F.; Pantelides, S. T.; Bolotin, K. I.; *Nano Lett.* **2013,** 13, 3626–3630.

(16) Desai, S. B.; Seol, G.; Kang, J. S.; Fang, H.; Battaglia, C.; Kapadia, R.; Ager, J. W.; Guo, J.; Javey, A. *Nano Lett.* **2014,** 14, 4592–4597.

(17) Ramasubramaniam, A.; Naveh, D.; Towe, E. *Phys. Rev. B* **2011,** 84, 205325.

(18) Zibouche, N.; Philipsen, P.; Kuc, A.; Heine, T. *Phys. Rev. B* **2014,** 90, 125440.

(19) Splendiani, A.; Sun, L.; Zhang, Y.; Li, T.; Kim, J.; Chim, C. Y.; Galli, G.; Wang, F. *Nano Lett.* **2010,** 10, 1271–1275.

(20) Mak, K. F.; Lee, C.; Hone, J.; Sahn, J.; Heinz, T. *Phys. Rev. Lett.* **2010,** 105, 136805.

(21) Ishigami, M.; Sau, J. D.; Aloni, S.; Cohen, M. L.; Zettl, A. *Phys. Rev. Lett.* **2005,** 94, 056804.

(22) Zheng, F.; Liu, Z.; Wu, J.; Duan, W.; Gu, B. *Phys. Rev. B* **2008,** 78, 085423.

(23) Dolui, K.; Pemmaraju, C. D.; Sanvito, S. *ACS Nano* **2012,** 6, 4823–4834.

(24) Riley, J. M.; Meevasana, W.; Bawden, L.; Asakawa, M.; Takayama, T.; Eknapakul, T.; Kim, T. K.; Hoesch, M.; Mo, S. K.; Takagi, H.; Sasagawa, T., Bahramy, M. S.; King, P. D. C. *Nat. Nanotechnol.* **2015,** 10, 1043–1047.

(25) Liu, H.; Han, N.; Zhao, J. *RCS Advances* **2015,** 5, 17572–17581.

(26) Amin, B.; Kaloni, T. P.; Schwingenschlogl, U. *RCS Advances* **2014,** 4, 34561–34565.

(27) Zeng, H.; Cui, X. *Chem. Soc. Rev.* **2015,** 44, 2629–2642.

(28) Wilson, N. R. Nguyen, P. V.; Selyer, K. L.; Rivera, P.; Marsden, A. J.; Laker, Z. P. L.; Constantinescu, G. C.; Kandyba, V.; Barinov, A.; Hine, N. D. M.; Xu, X.; Cobden, D. H. Preprint at http://arxiv.org/abs/1601.05865 **2016**.

(29) Brumme, T.; Calandra, M.; Mauri, F. *Phys. Rev. B* **2015,** 91, 155436.

(30) Cuong, N. T.; Otani, M.; Okada S. J. Phys.: Condens. Matter **2014,** 26, 135001.

(31) Podzorov, V.; Gershenson, M. E.; Kloc, C.; Zeis, R.; Bucher, E. *Appl. Phys. Lett.* **2004,** 84, 3301–3303.

(32) Eknapakul, T. *et al.* *Nano Lett.* **2014,** 14, 1312–1316.

(33) Lundgren, E.; Anderson, J. N.; Qvarford, M.; Nyholm, R. *Surf. Sci.* **1993,** 281, 83–90.

(34) Shanavas, K. V.; Satpathy, S. *Phys. Rev. B* **2015,** 91, 235145.